
\magnification = 1200
\def\lapp{\hbox{$ {     \lower.40ex\hbox{$<$}
    \atop \raise.20ex\hbox{$\sim$}
    }  $}  }
\def\rapp{\hbox{$ {     \lower.40ex\hbox{$>$}
    \atop \raise.20ex\hbox{$\sim$}
    }  $}  }
\def\barre#1{{\not\mathrel #1}}
\def\krig#1{\vbox{\ialign{\hfil##\hfil\crcr
    $\raise0.3pt\hbox{$\scriptstyle \circ$}$\crcr\noalign
    {\kern-0.02pt\nointerlineskip}
$\displaystyle{#1}$\crcr}}}
\def\upar#1{\vbox{\ialign{\hfil##\hfil\crcr
    $\raise0.3pt\hbox{$\scriptstyle \leftrightarrow$}$\crcr\noalign
    {\kern-0.02pt\nointerlineskip}
$\displaystyle{#1}$\crcr}}}
\def\ular#1{\vbox{\ialign{\hfil##\hfil\crcr
     $\raise0.3pt\hbox{$\scriptstyle \leftarrow$}$\crcr\noalign
     {\kern-0.02pt\nointerlineskip}
$\displaystyle{#1}$\crcr}}}

\def\g5{\gamma_5}

\topskip=0.60truein
\leftskip=0.18truein
\vsize=8.8truein
\hsize=6.5truein
\tolerance 10000
\hfuzz=20pt

\baselineskip 12pt plus 1pt minus 1pt
\pageno=0
\centerline{\bf SMALL MOMENTUM EVOLUTION OF THE EXTENDED}
\smallskip
\centerline{{\bf DRELL--HEARN--GERASIMOV SUM RULE
 }\footnote{*}{Work supported in part by Deutsche Forschungsgemeinschaft
and by Schweizerischer Nationalfonds.\smallskip}}
\vskip 24pt
\centerline{V\'{e}ronique Bernard}
\vskip 4pt
\centerline{\it Centre de Recherches Nucl\'{e}aires et Universit\'{e}
Louis Pasteur de Strasbourg}
\centerline{\it Physique Th\'{e}orique, BP 20 Cr, 67037 Strasbourg Cedex 2,
France}
\vskip 12pt
\centerline{and}
\vskip 12pt
\centerline{Norbert Kaiser}
\vskip 4pt
\centerline{\it Physik Department T30}
\centerline{\it Technische Universit\"at M\"unchen, James
Franck Stra{\ss}e}
\centerline{\it
D-8046 Garching, Germany}
\vskip 12pt
\centerline{and}
\vskip 12pt
\centerline{Ulf-G. Mei{\ss}ner\footnote{$^\dagger$}{Heisenberg
Fellow.}}
\vskip 4pt
\centerline{\it Universit\"at Bern}
\centerline{\it Institut f\"ur Theoretische Physik}
\centerline{\it Sidlerstr. 5, CH--3012 Bern,\ \ Switzerland}
\vskip 0.3in
\baselineskip 12pt plus 2pt minus 2pt
\centerline{\bf ABSTRACT}
\medskip
We investigate the momentum dependence of the extended Drell-Hearn-Gerasimov
sum rule. An economical formalism is developed which allows to express the
extended DHG sum rule in terms of a single virtual Compton amplitude in forward
direction. Rigorous results for the small momentum evolution are derived from
chiral perturbation theory within the one-loop
approximation. Furthermore, we
evaluate some higher order contributions arising from $\Delta(1232)$
intermediate states and  relativistic corrections.
\vfill
\noindent BUTP--92/51\hfill December 1992

\noindent CRN 92--53

\eject
\baselineskip 12pt plus 2pt minus 2pt
\noindent{\bf I. \quad INTRODUCTION}
\medskip
Many years ago, Drell and Hearn [1] and Gerasimov [2] (DHG) suggested a sum
rule for spin-dependent Compton scattering. It expresses the squared anomalous
magnetic
moment of the nucleon in terms of a dispersive integral over the difference of
the total photonucleon absorption cross sections $\sigma_{1/2}(\omega)$ and
$\sigma_{3/2}(\omega)$ for the scattering of circular polarized photons on
polarized nucleons. The subscripts $\lambda = 1/2$ and $\lambda =
3/2$ denote the total $\gamma N$ helicity, corresponding to states with photon
and nucleon spin antiparallel or parallel. Experimentally,
this sum rule has never been tested directly since up to now no measurements of
the helicity cross sections have been performed.
However, models for the photoabsorption  cross sections [3,4,5] do indicate its
approximate validity (on a qualitative level). One can now extend this sum rule
to virtual photons with $k^2 <0$ the  four-momentum transfer of the virtual
photon\footnote{*}{It is customary to set $k^2 = -Q^2$ and only use $Q^2$. We
will not do this in the following.} since the corresponding helicity cross
sections can be parametrized in terms of the spin-dependent nucleon structure
functions. The recent data of the European Muon
Collaboration [6] taken in the scaling region of large $|k^2| \simeq 10$
GeV$^2$ suggest not only that the pertinent sum rule behaves as
$1/k^2$ for large $|k^2|$, but
also that the sign is opposite to the DHG sum
rule for real photons (which in standard notations is negative). Therefore the
integral
$$I(k^2) = \int_{\omega_{thr}}^\infty {d\omega \over \omega} \bigl[
\sigma_{1/2}(\omega, k^2) - \sigma_{3/2}(\omega, k^2) \bigr] \eqno(1.1)$$
with $\omega$ the virtual photon energy in the nucleon rest frame must change
its sign between the photon point ($k^2 = 0$) and the EMC region, $k^2 \simeq -
10$ GeV$^2$. A recent
model predicts this turnover to happen at $k^2 \simeq -0.8$ GeV$^2$ [7] and it
explains this value mainly in terms of the low-energy contribution of the
$\Delta(1232)$ resonance to the pertinent photoabsorption cross sections.
Notice that the model of ref.[7] as well as the phenomenological analysis of
ref.[5] seem to indicate a negative slope of $I_p(k^2)$ in the vicinity of the
photon point, $k^2 \simeq 0$.

Here, we wish to add some new insight into the momentum dependence of the
integral $I(k^2)$ in the region of small $k^2$ where small means that $\sqrt
{- k^2} $ does not exceed a few pion masses. Our model--independent analysis is
based on the fact
that at low energies, the interactions of hadrons are governed by chiral
symmetry and gauge invariance (when external photons are involved). One can
systematically solve the chiral Ward-Takahashi identities of QCD via an
expansion in
external momenta and quark masses, which are considered small against the
scale of chiral symmetry breaking, $\Lambda_\chi \simeq 1$
GeV. This method is called chiral perturbation theory. It uses the framework of
an effective lagrangian of the asymptotically observed fields. The low-energy
expansion corresponds to an expansion in pion loops. In the presence of
baryons, a complication arises. The nucleon (baryon) mass in the chiral limit
is comparable to the chiral scale $\Lambda_\chi$ and thus only baryon
three-momenta can be considered small [8]. One can, however, restore the exact
one-to-one
correspondence between the loop and low-energy expansion using a
non-relativistic formulation of baryon chiral perturbation theory [9]. The
nucleon is considered as a very heavy (static)
source and in that case, all momenta involved are small therefore restoring the
consistent power counting. In what follows, we will use the non-relativistic
version of baryon CHPT which was systematically
investigated in ref.[10] as well as the relativistic formulation as spelled out
in detail in ref.[8].
This will allow us to extract the leading term in the chiral expansion of
$I(k^2)$ and to calculate the derivative of
$I(k^2)$ around $k^2 \simeq 0$. This is the region where CHPT applies.
Furthermore, following the suggestion of Jenkins and Manohar [11], we will also
add the $\Delta(1232)$ resonance to  non-relativistic baryon CHPT. The
$\Delta(1232)$ is the
lowest nucleon excitation and its closeness to the nucleon mass, $m_\Delta - m
\simeq 2.1\, M_\pi$, might indicate substantial contributions from it (this
is also
supported
by phenomenological models). In fact, using these various approximation
schemes, we will get a band of values for the slope of $I(k^2)$. Our most
important result, however, is that independent of the scheme we are using, we
find that $I(k^2)$ increases as $|k^2|$ increases (around $k^2 \simeq 0$). This
new result should serve as a constraint for all model builders and should
eventually be seen in refined phenomenological analyses or directly from the
data (when they will become available).

The paper is organized as follows. In section II, we spell out an economical
formalism to calculate $I(k^2)$ in terms of a single function which posseses a
right-handed cut starting at the single pion production threshold. This method
is considerably simpler than the one recently proposed by Meyer [12] whose
formalism involves half-off-shell nucleon form factors. In section III, we
use CHPT to
calculate $I(k^2)$ for the proton and the neutron at small $k^2$, in the
extreme non-relativistic and the fully relativistic formulation. The
contribution of
loops involving the $\Delta(1232)$ isobar in the non-relativistic approach
is also discussed.
The numerical results and conclusions are presented in
section IV.
\bigskip
\noindent{\bf II. \quad SPIN--DEPENDENT COMPTON SCATTERING: FORMALISM}
\medskip
In this section, we outline the formalism necessary to describe the scattering
of polarized (virtual) photons on polarized nucleons (protons and neutrons).
 Denote by $p$ and $k$ the four-momenta of
the nucleon and photon, respectively. It is convenient to work with the two
lorentz invariants $k^2$ and $\omega = p\cdot k /m$, with $m$ the nucleon mass.
The spin of the photon and nucleon can couple to the values $1/2$ and $3/2$
with the corresponding
photoabsorption cross sections
denoted by $\sigma_{1/2}(\omega,k^2) $ and
$\sigma_{3/2}(\omega, k^2)$, in order.\footnote{*}{For the definition of these
cross sections see ref.[13] (chap.2). We omit the tilde over the symbol
$\sigma$ used in that book.}
In what follows, we are interested in the extended
Drell-Hearn-Gerasimov sum rule, $i.e.$ the integral
$$I(k^2) = \int_{\omega_{thr}}^\infty {d\omega \over \omega} \bigl[
\sigma_{1/2}(\omega, k^2) - \sigma_{3/2}(\omega, k^2) \bigr] \eqno(2.1)$$
with $k^2 \le 0$ and the threshold photon energy $\omega_{thr}$ due to single
pion electroproduction is given by
$$\omega_{thr} = M_\pi + {M_\pi^2 - k^2 \over 2 m} \eqno(2.2)$$
where $M_\pi$ denotes the pion mass. For real photons, the expression (2.1)
becomes the celebrated DHG sum rule
$$I(0) = \int_{\omega_{thr}}^\infty {d\omega \over \omega} \bigl[
\sigma_{1/2}(\omega,0) - \sigma_{3/2}(\omega,0) \bigr] = - {\pi e^2 \kappa^2
\over 2 m^2 }\,. \eqno(2.3)$$
Here, $\kappa$ is the anomalous magnetic moment of the proton or the neutron
and we use standard units, $e^2 /4\pi = 1/137.036.$ The DHG sum rule is derived
under the assumption that the spin-dependent forward Compton amplitude for real
photons
$f_2(\omega^2) $ satisfies an unsubtracted dispersion relation which guarantees
that the right-hand side of eq.(2.3) converges. In what follows, we will make
use of the same assumption for virtual photons. To set the scale for $I(k^2)$,
let us give the numerical values for the proton and the neutron,
$$I_p(0) = -0.526\, {\rm GeV}^{-2}\,\,, \qquad I_n(0) = - 0.597\, {\rm
GeV}^{-2}\,. \eqno(2.4)$$
Our main concern will be the $k^2$ evolution of the extended DHG sum rule, in
particular around the origin $k^2 \simeq 0$. The interest in that comes from
the relation of the helicity cross sections to the spin-dependent nucleon
structure functions $G_1(\omega, k^2)$ and $G_2(\omega,
k^2)$. Following the notations of Ioffe {\it et al.} [13]\footnote{*}{We use a
different normalization  for the nucleon spinor, $\bar u u = 1$ instead of
$\bar u u = 2m$.}, one can show that
$$\sigma_{1/2}(\omega, k^2) - \sigma_{3/2}(\omega, k^2) = {4\pi e^2\over 2m
\omega + k^2 } {\omega \over m} \biggl[ G_1(\omega, k^2) + {k^2 \over m
\omega} G_2(\omega,k^2)  \biggr]\,. \eqno(2.5)$$
The relation of these structure functions to the spin-dependent virtual
Compton amplitudes in forward direction $S_{1,2}(\omega, k^2) $ is standard
$$2 \pi \, G_i(\omega, k^2 ) ={\rm Im} \, S_i(\omega, k^2) \,, \qquad (i = 1,2)
\eqno(2.6)$$
which follows from the optical theorem. Furthermore, crossing symmetry implies
that $S_1(\omega ,k^2)$ and $G_2(\omega ,k^2)$ are even functions under
($\omega \to - \omega) $ whereas $S_2(\omega,k^2) $ and $G_1(\omega,k^2)$ are
odd. In fact, for our purpose one does not need the information on both
amplitudes $S_1(\omega, k^2)$ and $S_2(\omega,k^2)$ but only the particular
combination entering eq.(2.5). In order to isolate this relevant combination
one contracts the antisymmetric (in $\mu \leftrightarrow \nu$) part of the
virtual Compton tensor in forward direction with polarization vectors
$\epsilon'_\mu$ and $\epsilon_\nu$ for the outgoing and incoming virtual
photon, respectively. If we choose the gauge conditions $\epsilon \cdot p =
\epsilon' \cdot p=\epsilon \cdot k= \epsilon' \cdot k = 0$ for the polarization
vectors and work in the nucleon rest-frame $p_\mu = (m,0,0,0)$ we obtain
$$\eqalign{
\epsilon'_\mu \, T^{\mu\nu}_{(a)} \, \epsilon_\nu &= {i\over 2 m^2}
\chi^\dagger \biggl\{ \vec \sigma \cdot (\vec\epsilon\,' \times \vec\epsilon\,)
\biggl[ \omega S_1(\omega,k^2) + {\omega^2 \over m} S_2(\omega, k^2) \biggr]
- \vec \sigma \cdot \vec k \, \vec k \cdot (\vec \epsilon\,' \times \vec
  \epsilon\,) {S_2(\omega,k^2) \over m} \biggr\}  \chi \cr
&= {i \omega\over 2m^2} \chi^\dagger \vec \sigma \cdot (\vec \epsilon\,' \times
\vec \epsilon\,) \chi \biggl[S_1(\omega, k^2) + {k^2 \over m \omega}
S_2(\omega, k^2) \biggr] \cr} \eqno(2.7)$$
where $\chi $ is a conventional two-component (Pauli) spinor. In eq.(2.7)
we have exploited the fact that under
the chosen gauge $\vec \epsilon\,'
\times \vec \epsilon$ is parallel to $\vec k$ and $\vec k\,^2 = \omega^2 -
k^2$.  Obviously, we are
projecting out the particular combination of $S_1(\omega,k^2)$ and
$S_2(\omega,k^2)$ whose imaginary part enters the extended DHG sum rule
$I(k^2)$. In analogy to the real photon case we call this combination
$$f_2(\omega^2, k^2) = {e^2 \over8 \pi m^2} \biggl[ S_1(\omega, k^2) + {k^2
\over m \omega} S_2(\omega, k^2)\biggr] \,.\eqno(2.8)$$
Here, we indicated already that $f_2(\omega^2,k^2) $ is an even function of
$\omega $ which follows from the $(\omega \to - \omega)$ crossing properties of
$S_{1,2}(\omega, k^2)$ [13]. The odd amplitude $\omega \, f_2(\omega^2, k^2)$
can now be expressed in terms of a single function $A(s,k^2)$ as follows
$$2\pi (s-m^2 - k^2 ) \, f_2(\omega^2, k^2) = e^2 \bigl[ A(s,k^2) - A(2m^2
+ 2 k^2 - s, k^2) \bigr]\,. \eqno(2.9)$$
Here, we introduced the Mandelstam variable $s
= (p + k)^2$ which is related to $\omega$ via $\omega = (s - m^2 - k^2 )/2m$.
The function $A(s,k^2)$ appearing in eq.(2.9) can always be chosen
such that it has only a right-handed cut starting at the single pion production
threshold $s = (m + M_\pi)^2$.
Under the assumption that $f_2(\omega^2, k^2)$
fulfills an unsubtracted dispersion relation (in $\omega$) or equivalently that
$A(s,k^2) $ fulfills a once-subtracted dispersion relation (in $s$, subtracted
at an arbitrary point $s_0$) we can make
use of the previous equations and calculate the extended DHG sum rule
$I(k^2)$ as
$$\eqalign{
I(k^2) &= 8\pi \int_{(m + M_\pi)^2 }^\infty ds {{\rm Im} \, f_2(\omega^2,k^2)
\over s - m^2}\cr &= 4 e^2 \int_{(m+ M_\pi)^2}^\infty ds { {\rm Im} A(s,k^2)
\over (s-m^2)(s - m^2 - k^2)} \cr
&= {4\pi e^2 \over k^2 } \biggl[ A(m^2 + k^2, k^2 ) - A(m^2, k^2) \biggr]
\, .\cr} \eqno(2.10)$$
This equation is our basic result. It is completely general and allows one to
calculate the extended DHG sum rule $I(k^2)$ from a single function $A(s,k^2)$
which can be easily computed from the virtual Compton tensor in forward
direction. To repeat it, eq.(2.10) was derived under the assumption that
$A(s,k^2)$ obeys a once-subtracted dispersion relation. That this is not a too
strong assumption $e.g.$ can be seen from the fact
that in the relativistic formulation of baryon CHPT to one-loop $A(s,k^2)$
indeed has this analytical property. However, a general proof for this is not
yet  available. In this sense the situation is analogous to $f_2(\omega^2,0)$
where the validity of an unsubtracted dispersion relation can not yet be proven
in general.  In the following section, we will use CHPT (in the one--loop
approximation) to evaluate $A(s,k^2)$ and to calculate $I(k^2)$ for $k^2$ in
the vicinity of zero (this is where CHPT applies).
\bigskip
\noindent{\bf III. \quad CHIRAL EXPANSION}
\medskip
At low energies, any QCD Green function can be systematically expanded   in
powers of small
momenta and quark (pion) masses. This is done within the framework of an
effective chiral lagrangian of the asymptotically observed fields, here the
nucleons, pions and photons. The low-energy expansion amounts to an expansion
in (pion) loops of the effective theory. In the presence of baryons, a
complication arises due to the baryon mass which is non-vanishing in the chiral
limit and therefore adds a new scale to the theory. In that case there is in
general no strict one-to-one correspondence between the low energy and loop
expansion. Stated differently, there is no guarantee that all next-to-leading
order
corrections at order $q^3$ (with $q$ denoting a generic small momentum) are
given completely by the one loop graphs. All calculations performed so far,
however, indicate that the leading non-analytic terms
(in the quark masses) which
arise due to infrared singularities in the chiral limit of vanishing pion mass
are indeed produced. Furthermore one also gets in the one loop approximation an
infinite  tower of higher order terms [8] which spoil the one-to-one mapping
between low-energy and loop expansion. To overcome these difficulties, it was
recently
proposed to use a heavy fermion effective field theory, $i.e.$ considering the
baryons as very heavy [9] and to expand the theory in inverse powers of the
baryon mass. In that case, the $n$-loop contributions are
suppressed by relative powers of $q^{2n}$ (with $q$ a genuine small
momentum) and a consistent counting scheme emerges. Furthermore, in this
framework one can easily couple in the $\Delta(1232)$ resonance since one
does not encounter the usual problems with the relativistic spin-3/2 particle
[11].
Nevertheless, we have to stress that the baryon mass $m$ comparable to the
chiral symmetry breaking scale $\Lambda_\chi$ is not very large. Therefore, an
expansion in powers of $M_\pi/m$ is a priori not to be expected to converge
very fast. Such $M_\pi/m$ suppressed contributions are partly resummed in the
relativistic approach. Of course the evaluation
of all $M_\pi/m$ corrections is
necessary to judge the quality of the chiral expansion.
Furthermore, once the spin--3/2 decuplet is included, one has an
extra non--vanishing scale in the chiral limit (the average
octet--decuplet mass splitting) which again complicates the low
energy structure.

The basic $\pi N \gamma$ lagrangian in the relativistic formulation of baryon
CHPT to leading order (${\cal O}(q)$) reads
$$\eqalign{
{\cal L} &= {\cal L}^{(1)}_{\pi N} + {\cal L}^{(2)}_{\pi \pi} \cr
 {\cal L}_{\pi N}^{(1)} &= \bar \Psi (i \barre D - m + {g_A\over 2}  \barre
 u \gamma_5 ) \Psi \cr
{\cal L}_{\pi \pi}^{(2)} &= {F^2\over 4} {\rm Tr} [ \nabla_\mu U
\nabla^\mu U^\dagger + M_\pi^2 (U+U^\dagger) ] \cr } \eqno(3.1)$$
where $U = \exp [i \vec \tau \cdot \vec \pi/ F] $ embodies the Goldstone
bosons, $u = \sqrt{U}$ and $u_\mu = i u^\dagger \nabla_\mu U u^\dagger$ with
$\nabla_\mu$ the pertinent covariant derivative. The isospinor $\Psi$ contains
the proton and neutron fields. The superscript $(i)$ denotes the chiral power
of the corresponding terms, it   counts derivatives and meson masses. The
construction of this effective lagrangian is unique. Let us point out that it
contains four parameters. These are the pion decay constant $F$, the
axial-vector coupling $g_A$ and the nucleon mass (all in the chiral limit) and
the leading term in the quark mass expansion of the pion mass, $M_\pi =
\sqrt{2\hat m B}$. Here, $\hat m = {1\over 2}(m_u + m_d)$ is the average light
quark mass and $B =- <0|\bar u u
|0>/F^2 $ is the order parameter of the spontaneous chiral symmetry breaking.
Calculating tree diagrams with this effective lagrangian, one reproduces the
well-known current algebra results. To restore unitarity, one has to consider
pion loops in additon. To give all corrections at next-to-leading order in the
chiral expansion one has to work
out all one loop diagrams constructed from the vertices in ${\cal L}$ and
furthermore one has to add the tree graph contribution from the most general
chirally symmetric counterterm lagrangian ${\cal L}^{(2)}_{\pi N} + {\cal
L}_{\pi N}^{(3)} + {\cal L}^{(4)}_{\pi \pi} $. For the (spin-dependent) Compton
 tensor under consideration here, however,
no such counterterm can contribute. As stressed in ref.[10], we are dealing
with a pure loop effect (within the one-loop approximation).

As already noted, in eq.(3.1) the troublesome nucleon mass term appears. In the
extreme non-relativistic limit, it can be eliminated in the following way.
Decompose the baryon four momentum as $p_\mu = m v_\mu + l_\mu$ with $v_\mu $
the four-velocity ($v^2 = 1$) and $l_\mu$ a
small off-shell momentum ($v\cdot l \ll m, \,$) and write $\Psi $ in terms of
eigenstates of the velocity projection operator
$$\Psi = e^{-im\,v\cdot x} ( H + h) \eqno(3.2)$$
with $\barre v H
= H$ and $\barre v h = - h$. Eliminating now the "small"
component $h$ via its equation of motion, one ends up with
$${\cal L}^{(1)}_{\pi N} = \bar H (i v \cdot D + g_A S \cdot u ) H +{\cal
O}(1/m)  \eqno(3.3)$$
Here, $S_\mu = {i\over2} \gamma_5 \sigma_{\mu\nu} v^\nu$ is the covariant spin
operator which obeys $S\cdot v = 0$. The nucleon mass term has disappeared
allowing for a consistent chiral power counting scheme. All one loop
contributions are order $q^3$. Furthermore, one has to expand the tree
contributions from the vertices of eq.(3.1) in $1/m$ appropriately to collect
all terms up to and including order $q^3$.
For a more detailed discussion of these topics, see ref.[10]. One can
furthermore add the $\Delta(1232)$, which is a spin-3/2 field, very easily in
the extreme non-relativistic limit. For details on the couplings of the
$\Delta(1232)$
see the appendix. Here, we just note that the mass splitting $m_\Delta - m$
stays finite in the chiral limit. Therefore loops with intermediate
$\Delta(1232)$ states will count as order $q^4$ and higher (since
the counterterm contributions start only at order $q^5$).

Let us now turn to the calculation of $I(k^2)$ for small $k^2$. In Fig.1.a we
show the pertinent Feynman diagrams which contribute in the heavy mass limit
(with intermediate nucleons only). We work in the Coulomb gauge $\epsilon'
\cdot v = \epsilon \cdot v = 0$ which is very economical in the calculation
of photon-nucleon processes since most diagrams (those with an isolated
photon-nucleon vertex) are then identical to zero. The integral $I(k^2)$ takes
the form
$$I(k^2) = I(0) + \tilde I(k^2) \eqno(3.4)$$
with $I(0) = - \pi e^2 \kappa^2 / 2m^2$ the DHG sum rule value for real
photons. In the heavy mass formulation of baryon CHPT the leading term of the
chiral expansion of $\tilde I(k^2)$ is given completely by the one loop
graphs in Fig.1a. All higher order corrections to $\tilde I(k^2)$  are
suppressed by further powers of the pion mass $M_\pi$ and $k^2$. Some (but not
all) of these corrections will be generated from loop diagrams with
$\Delta(1232)$ intermediate states or in the relativistic version of
baryon CHPT. The leading term of the chiral expansion of $\tilde I(k^2)$ can be
given in closed from
$$\tilde I(k^2) = {e^2 g_A^2 \over 4 \pi F^2} \biggl[ - 1 + \sqrt {1 + {4 \over
\rho}} \ln \biggl( \sqrt{1 + {\rho \over 4} } + {{\sqrt \rho} \over 2 } \biggr)
\biggr] = {e^2 g_A^2 \over 48 \pi F^2 } \, \rho + {\cal O}(\rho^2) \eqno(3.5)$$
with $\rho = - k^2 /M_\pi^2 > 0$. We see that the slope of $I(k^2)$ at $k^2 =
0$ is negative and singular in the chiral limit, $i.e$ it diverges like
$1/M_\pi^2$. This behaviour is a direct consequence of the chiral structure of
QCD which governs the low-energy strong interaction phenomena. Furthermore,
$\tilde I(k^2)$ is
equal for both proton and neutron (within the ${\cal O}(q^3)$ approximation to
the virtual Compton tensor). We should also add here that presently the usual
DHG sum rule value $I(0)$ for real photons can not be obtained through a
dispersive integral like eq.(2.10) within the
one-loop approximation of CHPT. In the heavy mass formulation this term arises
from real $1/m^2$ suppressed tree graphs involving the anomalous magnetic
moment $\kappa$ (in the chiral limit). In the relativistic version of baryon
CHPT the anomalous magnetic moment of the nucleon is generated from one loop
diagrams and it is non-vanishing in the chiral limit. In order to obtain a term
proportional to $\kappa^2$ like $I(0)$ one necessarily has to go to the level
of two-loop graphs. This problem of how $I(0)$ can be obtained from a
dispersion relation for loop amplitudes  does, however, not affect our
discussion of the
$k^2$ dependence of $I(k^2)$. Extending the effective lagrangian to the
$\Delta(1232)$ resonance as spelled out in the appendix we have to calculate
the diagrams of Fig.1b. These amount to some higher order ($q^n,{n\ge 1 }$)
corrections to eq.(3.5) which we include because of the phenomenological
importance of this resonance (a complete evaluation of all ${\cal O}(q)$
corrections to $I(k^2)$ corresponding to ${\cal O}(q^4)$ for the virtual
Compton tensor goes beyond the scope of this paper). A straightforward
calculation gives for the sum of nucleon and $\Delta(1232) $ one-loop diagrams
$$\eqalign{
\tilde I(k^2) =& {e^2 g_A^2 \over 4 \pi F^2} \biggl[ {r \over \sqrt{r^2 - 1}}
\ln \bigl( r + \sqrt{r^2 -1} \bigr) \cr &- \int_0^1 dx {r\over \sqrt {r^2 - 1
- \rho
x(1-x) } } \ln \biggl( {r \over \sqrt{1 + \rho x(1-x)}} + \sqrt{ {r^2 \over 1 +
\rho x(1-x) } - 1 }\biggr) \biggr] \cr }\eqno(3.6)$$
with $ r = (m_\Delta - m)/M_\pi \simeq 2.1$. Obviously, $\tilde I(0) = 0$ in
agreement with the celebrated low-energy theorem of Low, Gell-Mann and
Goldberger [15]. As a check one can show that in
the limit $m_\Delta - m \to \infty$ one recovers the result of eq.(3.5). Again
there is no splitting between proton and neutron sum rules, {\it i.e.} $ \tilde
I(k^2) = \tilde I_p(k^2) = \tilde I_n(k^2)$. The slope of the extended DHG sum
rule at the photon point is given as
$$I'(0) =  - {e^2 g_A^2 \over 48 \pi F^2 M_\pi^2 } { r^2 \sqrt{r^2 -1} - r \ln
(r + \sqrt{r^2 -1}) \over (r^2 - 1)^{3/2}} \,.\eqno(3.7)$$

In the relativistic formulation matters are different. First one has to
calculate many more Feynman diagrams. These generate some of the $M_\pi/m$
suppressed higher order corrections and naturally lead to a splitting between
proton and neutron for the momentum dependence of the extended DHG sum rule,
{\it i.e.} $\tilde I_p(k^2)\ne\tilde I_n(k^2)$.
What is conceptually most important is that in the relativistic version of
baryon CHPT one can indeed show that
the amplitude function $A(s,k^2)$ obeys a
once--subtracted
dispersion relation. Using now the definitions of the various loop functions as
given in ref.[14] extended to $k^2 \le 0$, the following expressions can be
deduced for $\tilde I_{p}(k^2)$ and $\tilde I_n(k^2)$
$$\eqalign{
\tilde I_p(k^2) =\,& {e^2 g_A^2 m^2\over 4\pi F^2 k^2} \int_0^1 dx \int_0^1 dy
\biggl\{ \bigl( 1 - {3m^2 \over k^2} y \bigr) \ln {M_\pi^2 (1-y) + m^2 y^2 +
k^2 y(y-1) \over M_\pi^2(1-y) + m^2 y^2} \cr
&- 2y  \ln {M_\pi^2 (1-y) + m^2 y^2 + k^2 xy(xy-1) \over M_\pi^2(1-y) + m^2 y^2
+ k^2 x(x-1) y^2} + {{3\over 2} y(1-y) [ k^2 - (2m^2 + k^2)y] \over M_\pi^2
(1-y) + m^2 y^2 } \cr &- {2k^2 xy(1-y)^2 \over M_\pi^2(1-y) + m^2 y^2
+ k^2 x(x-1) (1-y)^2} +  {y^2[k^2 (xy+ x^2y -{1\over2}) + m^2 (y-1) \over
M_\pi^2(1-y) + m^2 y^2+ k^2 x(x-1) y^2} \cr
&+{y^2[m^2(1-y) -k^2 x^2y]\over M_\pi^2(1-y) + m^2 y^2+ k^2 xy(xy-1)}
+ {2m^2k^2(1-x) xy^4 \over [M_\pi^2(1-y) + m^2 y^2 + k^2 xy(xy-1)]^2}
\cr &+ {y^4[k^4
x^2(1-x)(y-{1\over2} ) + m^2k^2({1\over 2} - {3\over 2} x + 2x^2 - xy) ] \over
 [M_\pi^2(1-y) + m^2 y^2+ k^2 x(x-1)y^2]^2} - {{3\over
2}k^2 m^2 y^3 (1-y)^2\over [M_\pi^2(1-y) + m^2 y^2]^2}
\biggr\}\cr} \eqno(3.8)$$
$$\eqalign{
\tilde I_n(k^2) =\,& {e^2 g_A^2 m^2\over 4\pi F^2 k^2} \int_0^1 dx \int_0^1 dy
\biggl\{ 2( 1 - y) \ln {M_\pi^2 (1-y) + m^2 y^2 +
k^2 x(y-1)(1-x+xy) \over M_\pi^2(1-y) + m^2 y^2 +k^2x(x-1)(1-y)^2} \cr
&+ 2y  \ln {M_\pi^2 (1-y) + m^2 y^2 + k^2 xy(xy-1) \over M_\pi^2(1-y) + m^2 y^2
+ k^2 x(x-1) y^2} + {y^2 [-2m^2 + k^2(2x-1)] \over M_\pi^2
(1-y) + m^2 y^2 + k^2x(x-1)y^2} \cr &+ {2m^2 y^2 \over M_\pi^2(1-y) + m^2 y^2
+ k^2 xy(xy-1)} +  {k^2(1-x)y^4[k^2x^2 + m^2 (1-4x)] \over
[M_\pi^2(1-y) + m^2 y^2+ k^2 x(x-1) y^2]^2} \cr &+
{4m^2k^2 x(1-x)y^4 \over [M_\pi^2(1-y) + m^2 y^2+ k^2 xy(xy-1)]^2} \biggr\}
\cr }\,.\eqno(3.9)$$
As an important analytical check we can again verify that $\tilde I_p(0) =
\tilde I_n(0) = 0$ and one can show that in the limit $m \to \infty$
both $\tilde I_p(k^2) $ and $\tilde I_n(k^2)$ tend to $\tilde I(k^2)$ as given
in eq.(3.5). With this we have collected all formulae necessary to study
$I(k^2)$ for both the proton and the neutron.
\bigskip
\noindent{\bf IV. \quad RESULTS AND DISCUSSION}
\medskip
First, we must fix parameters. Throughout, we use $F = 93$ MeV, $M_\pi =139.57$
MeV, $m = 938.27$ MeV and $g_A = 1.26$. In the case of the $\Delta(1232)$
resonance, we use the $SU(4)$ relation among coupling constants $g_{\pi N
\Delta} =
{3} g_{\pi N}/ \sqrt 2 $ with $g_{\pi N} = g_A \,{m/ F} $ given by the
Goldberger-Treiman relation. The mass splitting between nucleon and
$\Delta(1232)$ has a value of $m_\Delta - m = 293$ MeV.

Consider now the proton. We will first discuss the slope of $I_p(k^2)$ at the
photon point, $k^2 = 0$. In the
heavy mass limit with only intermediate nucleon states we find
$${d I_p(k^2)\over dk^2}\biggr|_{k^2 = 0} = - {e^2 g_A^2 \over 48 \pi F^2
M_\pi^2} =  - 5.7\,\,
{\rm GeV}^{-4}  \eqno(4.1)$$
This value is decreased by 16$\%$ when the $\Delta(1232)$ resonance is included
in the one loop graphs as inspection of eq.(3.7) reveals. Therefore the $\Delta
(1232)$ does not play a major role in determining the slope of $I_p(k^2)$ in
our approach. Much more drastic is the
effect of the relativistic $M_\pi/m$ suppressed terms. In the fully
relativistic calculation where many (but not all) of such terms are included we
find  $I_p'(0) =-2.2$ GeV$^{-4}$ for the proton and $I_n'(0)=-1.7$ GeV$^{-4}$
for the neutron. In Fig.2, we show $\tilde I_p(k^2)$ for $-k^2 \le 0.25 $
GeV$^2$. In the heavy mass limit half of the value of $I_p(0)$ (in magnitude)
is reached at $k^2 \simeq 0.06 $ GeV$^2$. The crossover where $I_p(k^2)$ goes
from negative to positive values takes place at $k^2 \simeq - 0.15 $ GeV$^2$.
This is a very low value compared to previous phenomenological analysis but
compared to the pion mass scale $M_\pi^2$ it is already quite large, $k^2
\simeq = -7.7 \, M_\pi^2$. Therefore one can no longer trust the one loop
approximation in that region of $k^2$ where the sign change of $I_p(k^2)$
takes place. Including some higher order chiral corrections through loops with
$\Delta(1232)$ resonances, the momentum dependence of $\tilde I_p(k^2)$ becomes
softer and the corresponding
numbers decrease by roughly 30$\%$. The zero of $I_p(k^2)$ is now shifted to a
higher value of $k^2 \simeq - 0.23 $ GeV$^2$. In the relativistic formulation
of CHPT where in addition to the leading terms also many higher order
corrections are included,
$\tilde I_p(k^2) $ is much smaller than in the case of
infinite nucleon mass. This phenomenon, that higher order relativistic
correction are quite large was also observed in previous calculations of the
nucleon electromagnetic polarizabilities [14]. However, since the $M_\pi/m$
corrections generated in the one--loop
approximation of relativistic baryon CHPT
are by no means     complete, one can not draw any conclusions about the
convergence of the chiral expansion at the moment.

In summary, we have presented a novel formalism to calculate the momentum
dependence of the extended DHG sum rule at finite $k^2 \le 0$. A single
amplitude function $A(s,k^2)$ which enters the spin-dependent virtual Compton
tensor in forward direction  is sufficient to evaluate $I(k^2)$, as long
as $A(s,k^2)$ fulfills a once--subtracted dispersion relation.
We have used
baryon chiral perturbation theory to investigate the behaviour of the extended
DHG sum rule $I(k^2)$ in the vicinity of $k^2 = 0 $. We could give a (rather
wide) range of values for the slope $I_p'(0)$. Eventually, this prediction will
be tested experimentally, at present we consider it as a constraint
following from the chiral structure of QCD which will be useful for
phenomenological analysis and model-building.
\goodbreak
\bigskip
\noindent{\bf APPENDIX: THE $\Delta(1232)$ IN THE HEAVY MASS FORMULATION}
\medskip
Here, we discuss briefly the description of the $\Delta(1232)$ resonance in
the heavy mass formulation following ref.[11].
To leading order (up to ${\cal O}(q)$) the relevant effective lagrangian reads
(we write down only those terms which are actually needed for our purpose)
$${\cal L}^{(1)}_{\pi N \Delta} = -i \bar T^{\mu a} \, v\cdot D^{ab} \,T^b_\mu
+ \delta m \, \bar T^{\mu a} T^a_\mu + {3g_A\over 2 \sqrt 2} ( \bar T^{\mu a}
u_\mu^a  H + \bar H u_\mu^a T^{\mu a} )\,. \eqno(A.1)$$
The Rarita-Schwinger spinor $T^a_\mu$ with $a$ an isospin index and $\mu$ a
Lorentz index incorporates the four charge states of the $\Delta(1232)$ as
follows
$$T^1_\mu = {1\over \sqrt 2} \left( \matrix { \Delta^{++} - \Delta^0/\sqrt 3\cr
\Delta^+ / \sqrt 3 - \Delta^- \cr} \right)_\mu , \quad
T^2_\mu = {i\over \sqrt 2} \left( \matrix { \Delta^{++} + \Delta^0/\sqrt 3\cr
\Delta^+ / \sqrt 3 + \Delta^- \cr} \right)_\mu , \quad
T^3_\mu = - \sqrt{{2\over 3}} \left( \matrix { \Delta^{+} \cr \Delta^0 \cr}
\right)_\mu\,. \eqno(A.2)$$
Furthermore in the heavy mass limit this field is subject to the constraint
$v_\mu T^{\mu a} = 0$.
In (A.1) $\delta m = m_\Delta - m$ stands for the mass splitting of nucleon and
$\Delta(1232)$ and  $u^a_\mu = {i\over 2} {\rm Tr}(\tau^a u^\dagger \nabla_\mu
U u^\dagger) = - \partial_\mu  \pi^a/ F - e \epsilon^{a3b} A_\mu \, \pi^b /F +
\dots $ gives rise to the chiral couplings of pions and photons to the
$N\Delta$ system. We already exploited the $SU(4)$ relation $g_{\pi N\Delta} =
3 g_{\pi N} /\sqrt 2$ with $g_{\pi N} = g_A m /F$ between the $\pi N \Delta$
and $\pi NN$ coupling constant. The empirical information on the $\Delta \to
\pi N $ decay width confirms that this relation holds very well within a few
percent. In the heavy mass limit the propagator of the $\Delta(1232)$ reads
$$P^{\mu \nu} = {i\over v\cdot l - \delta m} \bigl[ v^\mu v^\nu - g^{\mu\nu} -
{4\over 3} S^\mu S^\nu \bigr] \eqno(A.3)$$
where $S_\mu$ is the covariant spin operator of heavy mass approach satisfying
$v\cdot S = 0$. Let us finally remark that this formulation of $\Delta(1232)$
couplings is completely equivalent to the usual isobar model as discussed in
ref.[16] for the special choice $v_\mu = (1,0,0,0)$. This  corresponds to the
standard non-relativistic description.
\bigskip
\centerline{\bf ACKNOWLEDGEMENTS}
\medskip
\nobreak
The work of (UGM) was supported through funds provided by a Heisenberg
fellowship.
\vfill
\eject
\centerline{\bf REFERENCES}
\medskip
\item{1.}S.D. Drell and A.C. Hearn, {\it Phys. Rev. Lett.\/} {\bf 16}
(1966) 908.
\medskip
\item{2.}S.B. Gerasimov, {\it Sov. J. Nucl. Phys.\/} {\bf 2} (1966) 430.
\medskip
\item{3.}I. Karliner, {\it Phys. Rev.\/} {\bf D7} (1973) 2717.
\medskip
\item{4.}R.L. Workman and R.A. Arndt, {\it Phys. Rev.\/} {\bf D45} (1992) 1789.
\medskip
\item{5.}V. Burkert and Z. Li, CEBAF preprint, CEBAF--PR--92--017
\medskip
\item{6.}J. Ashman {\it et al.}, {\it Phys. Lett.\/} {\bf B206} (1988) 364.
\medskip
\item{7.}V. Burkert and B.L. Ioffe, CEBAF preprint, CEBAF--PR--92--018.
\medskip
\item{8.}J. Gasser, M.E. Sainio and A. ${\check {\rm S}}$varc,
{\it Nucl. Phys.\/}
 {\bf
B 307} (1988) 779.
\medskip
\item{9.}E. Jenkins and A.V. Manohar, {\it Phys. Lett.\/} {\bf B255} (1991)
558.
\medskip
\item{10.}V. Bernard, N. Kaiser, J. Kambor
and Ulf-G. Mei{\ss}ner, {\it Nucl. Phys.\/} {\bf B388} (1992) 315.
\medskip
\item{11.}E. Jenkins and A.V. Manohar, {\it Phys. Lett.\/} {\bf B259} (1991)
353.
\medskip
\item{12.}H. Meyer, University of Regensburg preprint, TPR--92--40.
\medskip
\item{13.}B.L. Ioffe, V.A. Khoze and L.N. Lipatov, "Hard processes", Vol.1,
North Holland, 1984.
\medskip
\item{14.}V. Bernard, N. Kaiser, and Ulf-G. Mei{\ss}ner, {\it Nucl. Phys.\/}
{\bf B373} (1992) 364.
\medskip
\item{15.} F.E. Low, {\it Phys. Rev.\/} {\bf 96} (1954) 1428;

M. Gell-Mann and M.L. Goldberger, {\it Phys. Rev.\/} {\bf 96} (1954) 1433.
\medskip
\item{16.}T. Ericson and W. Weise, "Pions and Nuclei", Clarendon Press, Oxford,
1988.
\bigskip
\nobreak
\centerline{\bf FIGURE CAPTIONS}
\smallskip
\item{Fig.1.} a) One loop diagrams contributing to the spin-dependent Compton
tensor in the heavy mass formulation of CHPT. Dashed lines denote pions.

\item{} b) One loop Compton graphs including the $\Delta(1232)$ resonance in
the heavy mass approach (denoted by a thick line).
\medskip
\item{Fig.2.} The momentum dependence of the extended DHG sum rule $\tilde
I_p(k^2)$. The solid line
gives the one--loop result in the heavy mass limit of
baryon CHPT. The dashed line
is obtained from one--loop graphs involving nucleons
as well as $\Delta(1232)$ resonances. The dashed--dotted line
gives the result of the
relativistic version of baryon CHPT to one loop.
\vfill
\eject
\end